\documentclass{article}

  \usepackage{microtype}
  \usepackage{graphicx}
  \usepackage{subfigure}
  \usepackage{booktabs} 
  \usepackage{caption}
  \usepackage[hyphenbreaks]{breakurl}
  \usepackage[hyphens]{url}

  \usepackage[accepted]{sysml2019}
  \usepackage{amsmath}
  \usepackage{listings}
  \usepackage{tabularx}
  \usepackage{multirow}
  \usepackage{array}
  \usepackage{mathtools}
  \DeclarePairedDelimiter{\ceil}{\lceil}{\rceil}

  \newcolumntype{M}[1]{>{\centering\arraybackslash}m{#1}}
  \newcolumntype{L}[1]{>{\raggedright}p{#1}}

  \usepackage{listings}           
\usepackage{courier}
\definecolor{vbgray}{gray}{0.9}
\definecolor{darkgreen}{RGB} {0, 100, 0}
\definecolor{darkred}{RGB} {255, 0, 0}
\definecolor{orange}{RGB} {255, 128, 0}

\definecolor{red}{RGB}{255,0,0}
\definecolor{vbgray}{gray}{0.9}
\definecolor{darkgreen}{RGB} {0, 100, 0}
\definecolor{darkred}{RGB} {255, 0, 0}
\definecolor{blue}{RGB} {0, 135, 255}
\definecolor{yellow}{RGB} {224, 173, 0}
\definecolor{codegreen}{RGB}{52,123,0}
\definecolor{codegray}{rgb}{0.6,0.5,0.5}
\definecolor{codepurple}{rgb}{0.58,0,0.82}
\definecolor{backcolour}{rgb}{0.95,0.95,0.95}
\definecolor{lightback}{rgb}{0.95,0.95,0.95}

\lstdefinelanguage{Spatial}{
  basicstyle=\fontsize{7}{7}\selectfont\ttfamily,frame=tlbr,framesep=4pt,framerule=0pt,
  tabsize=2,
  basewidth={0.55em, 0.4em},%
  numbers=left,
  showspaces=false,
  keywordstyle=\bfseries,
  breaklines=true,
  columns=fixed,
  firstnumber=auto,
  showstringspaces=false,
  escapechar=@,
  escapeinside={(*@}{@*)},
  morestring=[b]",
  morestring=[b]',
  morecomment=[l]{//},
  morecomment=[s]{/*}{*/},
  backgroundcolor=\color{backcolour},
  commentstyle=\color{codegreen}, 
  numberstyle=\tiny\color{codegray},
  stringstyle=\color{codepurple},
  keywordstyle=[2]\color{blue},
  keywords=[2]{val, def, type},
  keywordstyle=[3]\color{yellow}\bfseries,
  keywords=[3]{Float, Float8, Int, String, T, Void, Bit, Half, FltPt},
  keywordstyle=[4]\color{orange}\bfseries,
  keywords=[4]{Matrix, Array},
  keywordstyle=[5]\color{blue}\bfseries,
  keywords=[5]{StreamIn, StreamOut, DRAM, ArgIn, ArgOut, HostIO, RegFile, Reg, SRAM, SRAM1, SRAM2, FIFO, LIFO, LUT, LineBuffer},
  keywordstyle=[6]\bfseries,
  keywords=[6]{enq, deq, load, store, scatter, gather, :=, push, pop, peek},
  keywordstyle=[7]\color{magenta},
  keywords=[7]{until, par, by, value},
  keywordstyle=[8]\color{red}\bfseries,
  keywords=[8]{C0,C1,C2,C3,C4,C5,C6,C7,C8,C9,C10},
  keywordstyle=\color{magenta}\bfseries,
  morekeywords={Foreach,Reduce,MemReduce,MemFold,Fold,Accel,Stream,FSM,Sequential,if,else,Parallel,Pipe, DummyPipe}
}

  \sysmltitlerunning{Serving Recurrent Neural Networks Efficiently with a Spatial Accelerator}

\begin{document}

  \twocolumn[
  \sysmltitle{Serving Recurrent Neural Networks Efficiently with a Spatial Accelerator}



  \sysmlsetsymbol{equal}{*}

  \begin{sysmlauthorlist}
  \sysmlauthor{Tian Zhao}{sf}
  \sysmlauthor{Yaqi Zhang}{sf}
  \sysmlauthor{Kunle Olukotun}{sf}

  \end{sysmlauthorlist}

  \sysmlaffiliation{sf}{Department of Electrical Engineering, Stanford University, Stanford, USA}

  \sysmlcorrespondingauthor{Tian Zhao}{tianzhao@stanford.edu}
  \sysmlcorrespondingauthor{Yaqi Zhang}{yaqiz@stanford.edu}
  \sysmlcorrespondingauthor{Kunle Olukotun}{kunle@stanford.edu}

  \sysmlkeywords{Model Serving, Parallel System}

  \vskip 0.3in
  \begin{abstract}
Recurrent Neural Network (RNN) applications form a major class of
    AI-powered, low-latency data center workloads.
Most execution models for RNN acceleration
    break computation graphs into BLAS kernels,
    which lead to significant inter-kernel data movement
    and resource underutilization.
We show that by supporting more general loop constructs that capture design parameters in accelerators,
it is possible to improve resource utilization using cross-kernel optimization without sacrificing programmability.
Such abstraction level enables a design space search that can lead to efficient usage of
  on-chip resources on a spatial architecture across a range of problem sizes.
We evaluate our optimization strategy on such abstraction with DeepBench using a configurable spatial accelerator.
We demonstrate that this implementation provides
    a geometric speedup of 30x in performance, 1.6x in area, and 2x in power efficiency
compared to a Tesla V100 GPU, and a geometric speedup of 2x
compared to Microsoft Brainwave implementation on a Stratix 10 FPGA.

\end{abstract}

  ]


  \printAffiliationsAndNotice{}

\section{Introduction}
\label{sec:intro}

Recurrent Neural Networks (RNNs) are a class of sequence models that plays a key role
  in low-latency,
  AI-powered services in datacenters \cite{fowers2018configurable, jouppi2017datacenter}.
In these services,
  the platforms assume that user requests come in individual samples
  and need to be served with very stringent latency window
  for real-time human computer interaction.
  An example of such workload is
  Google Translate, where inference happens concurrently when a user types.
Despite its popularity, RNN model serving is hard to accelerate efficiently.
Modern software and hardware platforms support optimized BLAS routines.
To serve RNNs on these platforms,
  a compiler tends to stitch multiple optimized BLAS kernels into a single computation graph.
While a hardware accelerator might execute each individual kernel efficiently,
it misses the opportunity of global cross-kernel optimization that can dramatically
improves performance and energy-efficiency.
This approach leads to two issues.
First, communication between BLAS kernels creates large intermediate results, which can
lead to poor memory performance when the blocking size is not properly tuned for the target
system.
Missing the opportunity of cross-kernel fusion can lead to huge performance loss
due to different access latency at each level of memory hierarchy in a processor-based
architecture. On a spatial architecture, while the first two levels of memory hierarchies,
  i.e. registers and on-chip scratchpads,
  tend to have single cycle access latency,
  the energy required to access these two types of memory would be widely different.
Therefore, lack of cross-kernel fusion can lead to inefficient allocation of
scratchpad resource and low energy-efficiency.
Second, hardware accelerators tend to use large vectorization in
compute and memory access to boost compute density
when accelerating BLAS kernels. However, hardware accelerators tend to suffer from resource
underutilization when the workload size is not multiples of the vector size.
The utilization is worse with RNN applications that are composed of sequences of small matrix
  multiplications due to small hidden unit sizes and many time steps.
Moreover, many accelerator platforms are optimized for BLAS level-3 (matrix-matrix) operations, e.g. NVBLAS Library for GPU \cite{nvblas}, TPU \cite{jouppi2017datacenter}, EIE \cite{han2016eie}, EyeRiss \cite{chen2017eyeriss}, and DaDianNao \cite{chen2014dadiannao}.
These platforms suffer from low resource utilization when serving single-batch,
  real-time RNN applications
  that contain a lot of matrix-vector multiplication (MVM) executions.

To address these issues, we propose the following strategies.
First, we fuse all the gate functions with the element-wise, non-linear functions in the same time step.
This way, all of our intermediate results are buffered in the registers as opposed to the
scratchpads.
Second, we spatially parallelize and pipeline the computation graph.
We vectorize the inner-loop of the tiled dot product to explore SIMD parallelism and fine-grain pipelining.
We also explore tiled parallelism and coarse-grain pipelining by unrolling the outer loop nests
based on the amount of available compute resources.
These strategies exploit the gate-level parallelism in RNN cells, balance the pipelines of
MVM and element-wise non-linear functions, and maximize the resource utilization when serving
RNN models on different problem sizes. In addition, the entire pipeline is data-flow
driven with no dynamic scheduling overhead.

We evaluate the proposed strategies by serving RNN tasks in DeepBench \cite{deepbench}
  on the target spatial architecture.
We implement the designs in Spatial \cite{spatial_koeplinger},
  a Domain-Specific-Language (DSL) that describes
  applications with nested loops and explicit
  hardware memory hierarchy.
We choose Plasticine \cite{plasticine},
  a coarse-grained reconfigurable architecture (CGRA),
  as the target spatial architecture.
Furthermore, we propose augmentations to the Plasticine microarchitecture
  in order to support the mix-precision operations,
  which is critical for serving RNNs in real-time.

Finally, we compare the results to those obtained by serving DeepBench tasks on the
  state-of-the-art RNN serving platforms.
We show that our implementation delivers consistently high FLOPS utilization across
tasks of various sizes.
We also demonstrate energy-efficiency advantage of
spatial architectures compared to processor-based architectures.

The key contributions of this paper are:
\begin{enumerate}
\item
  We analyze the computation and memory layout of
  RNN cell implementations on commercially available platforms.
  We find that BLAS abstraction leads to
  expensive inter-kernel data movement and resource underutilization.
\item
  We address these issues by describing RNN applications using
  abstractions with more general loop constructs that enable 
  cross-kernel optimization, spatial parallelization, and pipelining
  of arbitrary loop nesting.
  To achieve low-latency inference for RNN applications,
  we propose micro-architectural co-design to a spatial architecture in order
  to enable low-precision operations.
\item
  Finally, we thoroughly evaluate CPU,
    general purpose graphics processing unit (GPGPU),
    field-programmable gate array (FPGA), and a previously-proposed CGRA,
    as serving platforms for RNN applications.
\end{enumerate}

The rest of the paper is organized as follows.
Section~\ref{sec:back}
  provides backgrounds on the RNN algorithms,
  the DSL and hardware platform used in this paper.
Section~\ref{sec:app} discusses the available RNN implementations
  on commercially available platforms.
  We then discuss the optimization strategies implemented in this work
  that address the inefficiency in these implementations.
Section~\ref{sec:arch} discusses the architectural changes
  for supporting efficient RNN inference on the target spatial architecture.
Section~\ref{sec:eval} details our evaluation methodology and experimental results.
Section~\ref{sec:related} discusses related works on available
  software and hardware optimization strategies for serving RNN applications.
Section~\ref{sec:conclusion} offers concluding remarks.

  \section{Background}
\label{sec:back}
RNNs are widely used to model arbitrary sequential tasks.
An RNN contains a cell unit to iteratively consume
  a T-step input sequence $x = [x_0, x_1, \cdots, x_T]$
  in order to generate an output sequence $y = [y_0, y_1, \cdots, y_T]$.
Long Short-Term Memory (LSTM) \cite{hochreiter1997long}
  and Gated Recurrent Unit (GRU) \cite{chung2014empirical}
  are popular RNN cell units.
In this paper, we use LSTM as an example.
Nevertheless, our optimization techniques can be generalized to any other types of RNN cells.
In Section \ref{sec:eval},
  we also provide evaluations of GRU implemented using our techniques.

\subsection{LSTM Cell}
At step $t$, an LSTM generates an
  output $y_t$ and the next memory cell states $c_t$ and $h_t$ as follows:
  \begin{align}
    \centering
    i_t &= \sigma (W_{h_i} h_{t-1} + W_{x_i} x_t + b_i) \label{eq:1} \\
    j_t &= \tanh (W_{h_j} h_{t-1} + W_{x_j} x_t + b_j) \label{eq:2} \\
    f_t &= \sigma (W_{h_f} h_{t-1} + W_{x_f} x_t + b_f) \label{eq:3} \\
    o_t &= \sigma (W_{h_o} h_{t-1} + W_{x_o} x_t + b_o)\label{eq:4} \\
    c_t &= f_t \circ c_{t-1} + i_t \circ j_t \label{eq:5} \\
    y_t &= h_t = o_t \circ \tanh (c_t) \label{eq:6}
  \end{align}
$H, D$ are dimensions of hidden states and input features, respectively.
$R$ is the sum of hidden state and input feature dimensions.
$\circ$ is the Hadamard product.
Table \ref{tab:spec_lstm} shows the specifications for each matrix and vector in an LSTM cell.

\begin{table}[t]
\vskip 0.15in
\centering
\scriptsize
\begin{tabular}{p{1cm}m{1cm}m{5cm}}
\toprule
  Name & Shape & Specification \\
  \midrule
  $x_t$    & $D$      &  LSTM cell's input vector\\
  $f_t$    & $H$      &  Forget gate's activation vector\\
  $i_t$    & $H$      &  Input gate's activation vector\\
  $o_t$    & $H$      &  Output gate's activation vector\\
  $j_t$    & $H$      &  Candidate of memory gate's activation vector\\
  $c_t$    & $H$      &  Memory gate's vector \\
  $W_{h_{i,j,f,o}}$   & $H,H$    &  Hidden state's weight matrices at gate $i,j,f,o$\\
  $W_{x_{i,j,f,o}}$   & $H,D$    &  Input vector's weight matrices at gate $i,j,f,o$\\
  $b$      & $H$      &  Bias vector at gate $i$,$j$,$f$,$o$\\
\bottomrule
\end{tabular}
\caption{LSTM specifications.}
\label{tab:spec_lstm}
\vskip -0.1in
\end{table}

\subsection{Spatial Reconfigurable Architectures}
Spatial reconfigurable architectures, such as FPGAs and CGRAs, are gaining traction as data center accelerators for their
energy efficiency \cite{awsf1, catapult, baidu}.
Compared to processor-based architectures, spatial architectures can reach high resource utilization
  by reconfiguring memory and compute based on the applications and computation requirements.
In addition to exploiting parallelism, pipelining of data-flow graph in
  a spatial architecture provides high compute throughput.
Nonetheless, the traditional low-level programming
  interface and long synthesis time of FPGA is the major obstacle
  for it to become a mainstream accelerator.
As opposed to bit-level flat-interconnection in FPGAs,
  CGRAs are usually configured at higher level of granularity and contain a hierarchical interconnection network.
In exchange, the reduction in flexibility
  in hardware translates to lowered routing overhead and
  higher clock frequency.
The reduced routing
  overhead provides higher compute density and memory capacity,
  which makes CGRA an attractive
  platform to accelerate deep learning workloads.
Due to the flexibility in mapping applications,
  spatial architectures often require design space exploration (DSE)
  in order to achieve good resource
  utilization and performance \cite{dse_koeplinger, fpgadse}.

\subsection{Spatial}
Spatial is a hardware-centric DSL that targets FPGAs and a previously
  proposed CGRA, Plasticine.
A user describes applications in un-parallelized pattern-based loops with explicit
memory hierarchies.
Spatial automatically schedules, parallelizes, and pipelines arbitrary loop nests.
To scale the memory bandwidth with parallelism, Spatial banks the scratchpad memories.
To sustain the throughput of pipelining, Spatial also buffers the intermediate memories.
Spatial exposes important design parameters such as blocking size and unrolling factor.
Using the exposed parameters,
  users can easily tune their design either manually or with an external DSE engine to
  balance the pipeline stages and saturate resource for different tasks on different
  hardware targets.

\subsection{Plasticine}
Plasticine is a CGRA that accelerates general
  nested loops in Spatial.
It consists of primarily two types of units:
  a pattern compute unit (PCU) containing a single instruction multiple data (SIMD) pipeline
  optimized for accelerating vectorized map and reduction loops,
  and a pattern memory unit (PMU) containing configurable
  memory that to support banking and buffering schemes for various access patterns.
Plasticine supports parallelizing and pipelining arbitrarily nested loops from Spatial.
More architectural details will be explained in Section \ref{sec:arch}.

  \section{RNN Computation Analysis} 
\label{sec:app}
In this section, we first discuss the limitation of BLAS-based LSTM on processor and spatial architectures.
Next, we discuss our implementation of loop-based LSTM on spatial architectures.
Table \ref{tab:legend_app} contains specifications for symbols and parameters
  used in this section.
\begin{table}[t]
  \vskip 0.15in
  \centering
  \scriptsize
  \begin{tabular}{p{0.6cm}m{3cm}m{3cm}}
  \toprule
    Symbol & Processor & Reconfigurable Hardware \\
    \midrule
    \includegraphics[width=0.03\columnwidth]{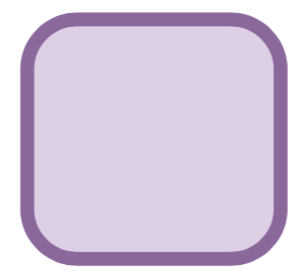} & Kernel & Inner Loop \\
    \includegraphics[width=0.03\columnwidth]{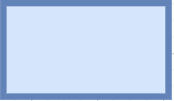} & Memory Hiearchy & On-chip Scratchpad \\
    \includegraphics[width=0.03\columnwidth]{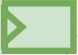} & Register File & Register \\
    \includegraphics[width=0.1\columnwidth]{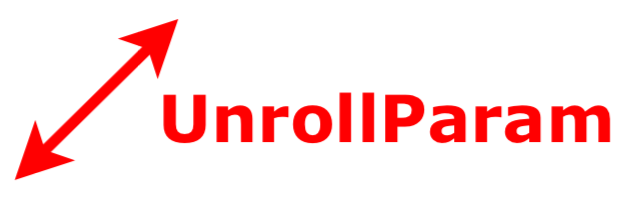} & & Unrolling factor using multiple hardware compute blocks \\ \midrule
    \includegraphics[width=0.03\columnwidth]{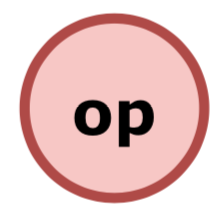} & \multicolumn{2}{L{6.5cm}}{Element-wise Operation} \\
    \includegraphics[width=0.03\columnwidth]{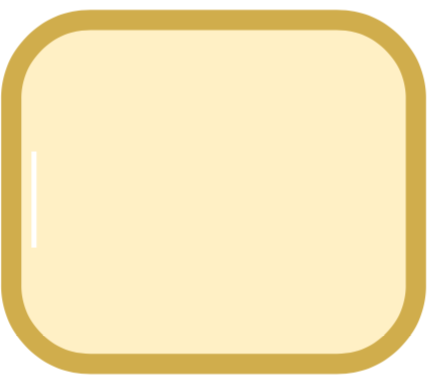} & \multicolumn{2}{L{6.5cm}}{Outer Loop} \\
    \includegraphics[width=0.1\columnwidth]{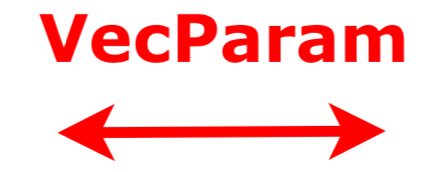} & \multicolumn{2}{L{6.5cm}}{Vectorization parameter for AVX or SIMD instructions} \\
    \midrule
    \midrule
    Parameter & \multicolumn{2}{l}{Specification} \\
    \midrule
    $hv$      & \multicolumn{2}{l}{Vectorization parameter on H} \\
    $hu$      & \multicolumn{2}{l}{Unrolling factor on H} \\
    $rv$      & \multicolumn{2}{l}{Vectorization parameter on R} \\
    $ru$      & \multicolumn{2}{l}{Unrolling factor on R} \\
    $G$       & \multicolumn{2}{l}{Number of gates in an RNN. For LSTM, G=4} \\
  \bottomrule
  \end{tabular}
  \caption{Specifications for symbols and parameters in Section \ref{sec:app}.}
  \label{tab:legend_app}
  \vskip -0.1in
  \end{table}

\begin{figure*}
  \centering
  \includegraphics[width=1.5\columnwidth]{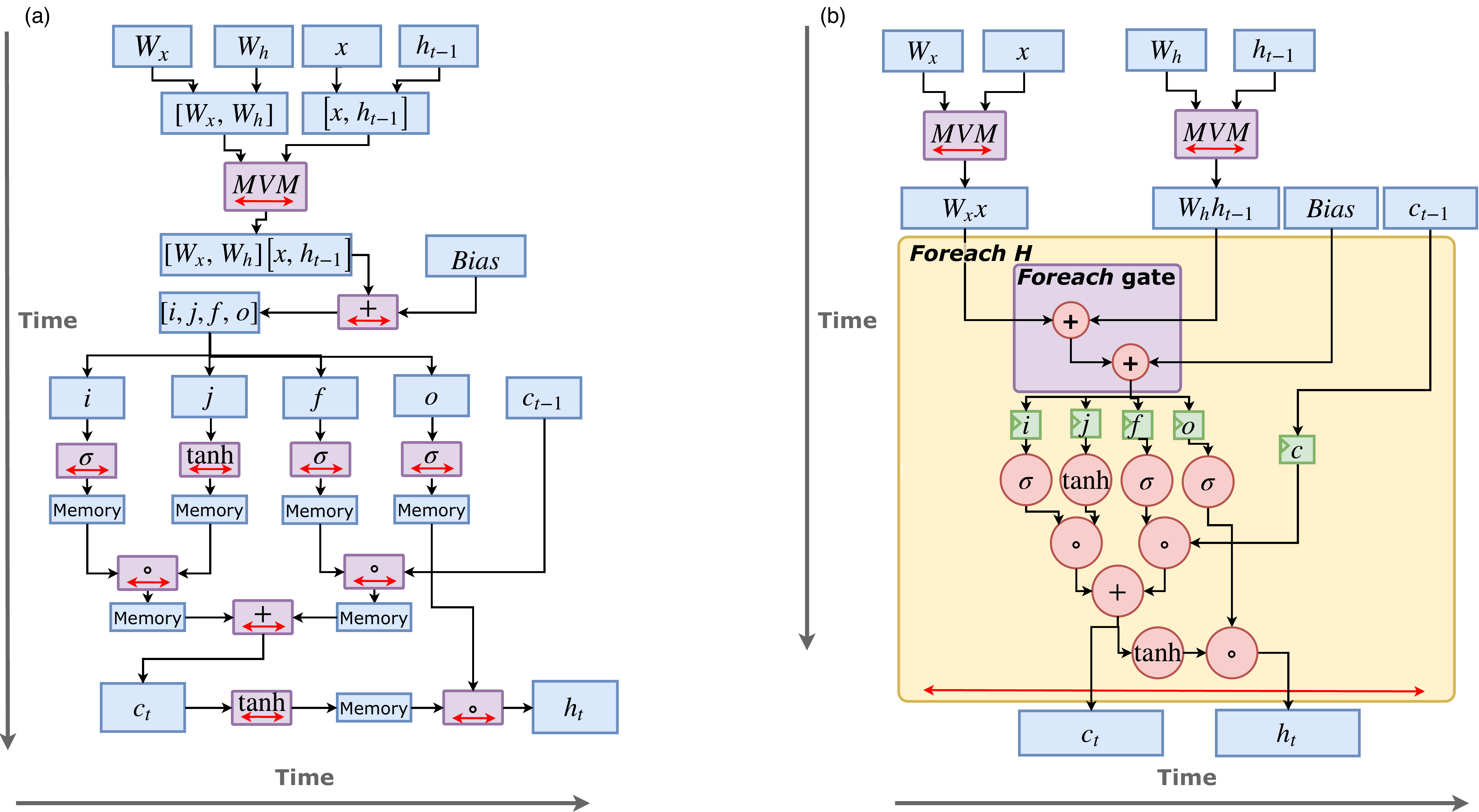}
  \caption{Compute and memory layout of TensorFlow \texttt{BasicLSTM} cell on CPU (a) and \texttt{CudnnLSTM} cell on GPU (b).}\label{fig:tf_lstm}
\end{figure*}

\subsection{BLAS-based LSTM on Processor Architecture}
Modern Machine Learning frameworks, e.g.
  TensorFlow \cite{abadi2016tensorflow},
  divide the computation graph of an LSTM cell into BLAS kernels.
Then, the BLAS kernel is accelerated by calling low-level
optimized BLAS subroutines such as Intel BLAS Library on CPU
and NVBLAS Library on GPU.
Figure \ref{fig:tf_lstm} (a) shows the computation graph of a \texttt{BasicLSTM} cell in TensorFlow.
This implementation can lead to large memory footprint since all the intermediate results are
  materialized in memory.
A common strategy to tackle the issue is through fusing blocked kernels.
With TensorFlow's abstraction, this can only be achieved by
  expressing the entire RNN cell as an optimized kernel.
For example, TensorFlow provides \texttt{LSTMBlockFusedCell} and \texttt{GRUBlockCell} modules,
  which are the fastest TensorFlow implementations of RNN cells for CPU.
In practice, such implementation can provide significant performance improvement
  over the \texttt{BasicLSTM} implementation.
However, it is still very hard to saturate CPU compute capacity, potentially
due to the high synchronization overhead across threads.
Figure \ref{fig:tf_lstm} (b) shows the computation layout of TensorFlow with
cuDNN library \cite{chetlur2014cudnn} on GPU. cuDNN is an NVIDIA GPU
library for accelerating deep neural networks.
To minimize the data movement,
  cuDNN fuses all the vector-vector (VV) operations after MVM. Specifically, the bias add in
  Equation \ref{eq:1}, \ref{eq:2}, \ref{eq:3}, \ref{eq:4},
  and all the operations in Equation \ref{eq:5}, \ref{eq:6},
  are fused into one kernel.
Nevertheless, there are still intermediate buffers of size $H$
  between the MVM kernel and the element-wise operations.

Compared to the \texttt{BasicLSTM} implementation,
  \texttt{CudnnLSTM} eliminates most of large intermediate memories.
However, the MVMs of Equation \ref{eq:1}, \ref{eq:2}, \ref{eq:3}, \ref{eq:4} are all accelerated
with BLAS3 kernels, which performs only matrix-matrix level operations.
This turns MVM and VV bias add into Matrix Matrix Multiplication (MMM) and Matrix Matrix
Addition (MMA), which leads to serious underutilization of GPU.

Moreover, a processor-based architecture introduces large energy overhead of instruction
  decoding and scheduling.
GPU especially suffers from its power-hungry, high-throughput memory hierarchy.
For these reasons, both the CPU and GPU architectures are not suitable
  for energy-efficient, low-latency RNNs serving platforms.

\subsection{BLAS-based LSTM on Spatial Architecture}

Previous work has studied the capability of using an FPGA as a low-latency serving platform.
An FPGA has the flexibility of resizing MVM and VV units based on the application size.
In addition, MVM and VV units can be implemented with hardware pipelines,
  which removes the instruction scheduling and control overhead on a processor-based
  architecture.
The latest version of Intel Stratix 10 FPGA further boosts the compute power of FPGA
  with increasing number of hardened digital signal processing (DSP) blocks
  and on-chip memory capacity.
Microsoft Brainwave (BW) \cite{fowers2018configurable}
  is a state-of-the-art FPGA-based deep learning framework.

\begin{figure}
  \centering
  \includegraphics[width=0.8\columnwidth]{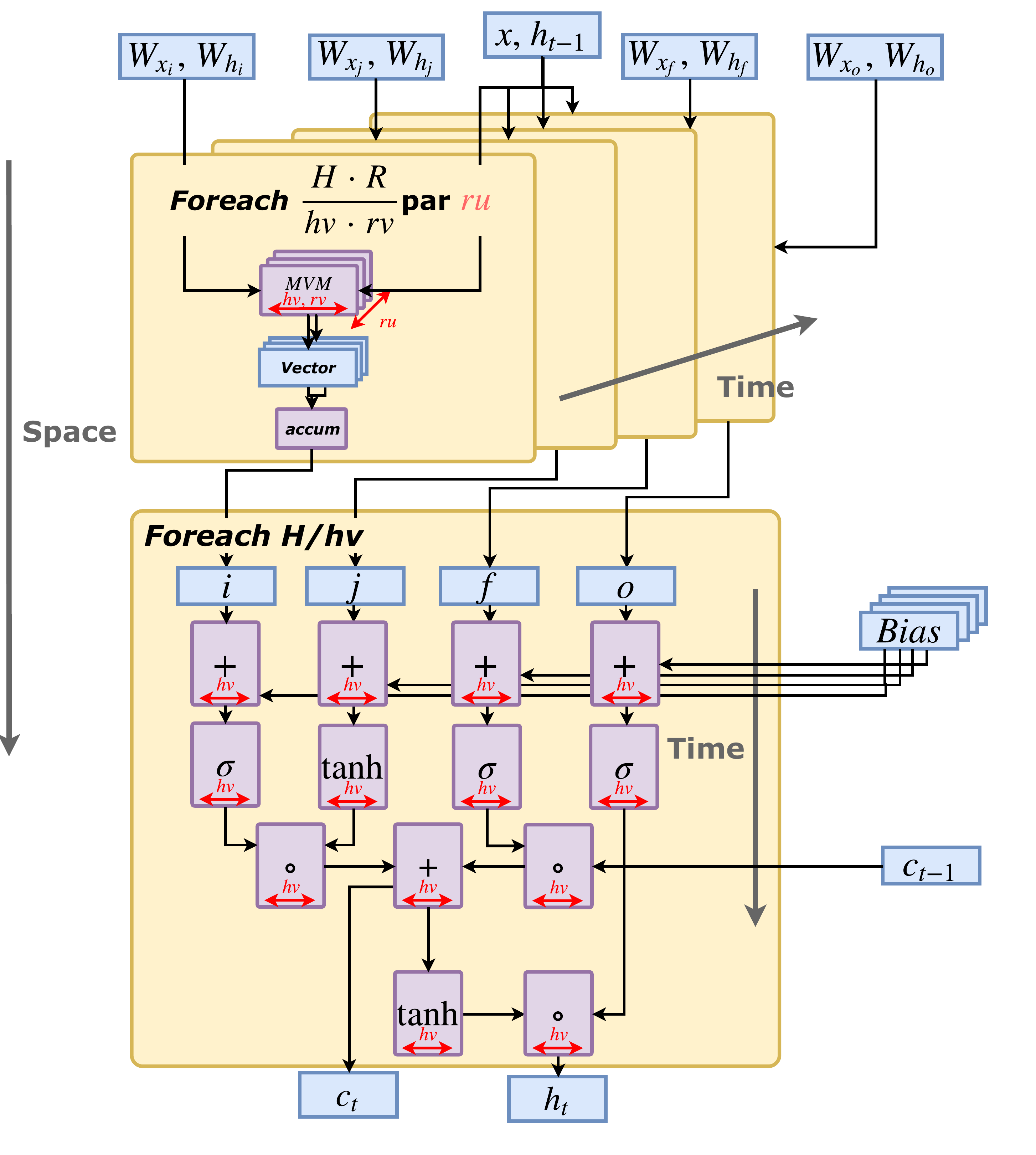}
  \caption{Compute and memory layout of LSTM in Brainwave.}\label{fig:bw_lstm}
   \vspace*{-0.3in}
\end{figure}

Figure \ref{fig:bw_lstm} shows BW's compute and memory layout.
In contrast to the CPU and GPU implementations, BW blocks the MVM along both
  row and column dimensions.
It then fuses the inner tiled MVM with element-wise non-linear functions.
Specifically for a matrix of size $H\times R$ and a vector of size $R$,
BW parallelizes the compute of multiple column tiles ($ru$, \# MV Tiles in the original paper) of size
$hv\times rv$ with multiple tiled engines, as shown in Figure \ref{fig:bwt} (a). 
Each tile engine contains $hv$ (native dimension) number of dot
product engines vectorized by $rv$ (lanes) and achieves one tile per cycle throughput. 
Parallel tiles along the row dimension are then fed into a pipelined reduction and accumulation
unit.
Immediately after the accumulation, the multi-function units (MFUs) execute the
element-wise operations on the $hv$ vector chunk produced by the accumulator.
Although BW's implementation still keeps the vectorized intermediate results, the size $hv$ is much
smaller than $H$ in \texttt{BasicLSTM} cell.
Nonetheless, with parallelization in $ru$,
  BW allocates lots of vectorized intermediate buffers that can still lead to energy inefficiency.
BW performs one MVM operation in $\ceil[\big]{\frac{H}{hv}}\ceil[\big]{\frac{R}{rv\cdot ru}}$
  iterations.

The MVM operations are executed on each gate of the LSTM sequentially.
Similarly, element-wise operations $hv$ using $\sigma, \tanh, \circ, +$ for the non-linear 
operators are also scheduled to execute on the vectorized multi-function units with size
of $hv$, as shown with the arrow in time in Figure \ref{fig:bw_lstm}.
To avoid DRAM communication overhead and improve compute density, Brainwave embeds MVM in a blocked floating-point format, 
where the vector of $hv$ values share a single 5-bit exponent and have distinct signs and 2-5 bit mantissa for
each value. As a result, they can achieve very dense low-precision compute and storage, with one
adder per $hv$ values and $hv$ multipliers for a vector of $hv$. The remaining operations are
performed in 16-bit precision.

When matrix dimensions cannot be divided by $hv$ and $rv\cdot ru$, Brainwave suffers from
underutilization of the compute FLOPS, as shown in Figure \ref{fig:bwt} (a).
The underutilization is worse with small problem sizes.
In addition, BW computes $W_xX$ and $W_hH$ separately rather than computing them with concatenated larger matrices,
  which can further aggravate the problem.
This might be because BW's abstraction does not allow partial updates of an vector
  but only $X$ is updated at the end of the step.

\subsection{Loop-based LSTM}
We have made the following observations
  from analyzing BLAS-based LSTM implementations:
\begin{enumerate}
\item 
  Constructing an LSTM cell's computation graph using BLAS subroutines
  introduces large intermediate buffers even when the kernels themselves are blocked.
  Each element on RNN cells' non-reduction dimension of the MVM ($H$)
  can be computed completely independently within one time step. This
  exposes the opportunity of fine-grain loop tiling and fusion across the
  entire LSTM kernel. 
\item MVM is the computation bottleneck in serving RNN cells. Spatial
  architecture allows us to distribute most of the compute resource to MVM 
  by parallelizing and pipelining MVM with element-wise operations.
\item Using low-precision operations can boost compute density and keep RNN
  weights on-chip to avoid high-latency DRAM communication. We need to introduce
  efficient low-precision support in the target spatial architecture without introduce
  too much overhead.
\end{enumerate}

To address the issue of large intermediate buffers, we fine-grain tile and fuse
MVM with non-linear functions. We refer to the computation for generating 
every single element in $c_t$ and $h_t$ as LSTM-1 operation, which can be computed
independently in a single step. LSTM-1 is composed of four independent dot products of 
the row vectors of the weight matrices with the input vector immediately followed by the 
element-wise operations on output of the dot product. The resulting $c$ and $t$ vectors are 
produced by computing LSTM-1 operations for $H+D$ iterations.

\label{sec:blaslstm}

\begin{figure}
 \centering
  \includegraphics[width=0.9\columnwidth]{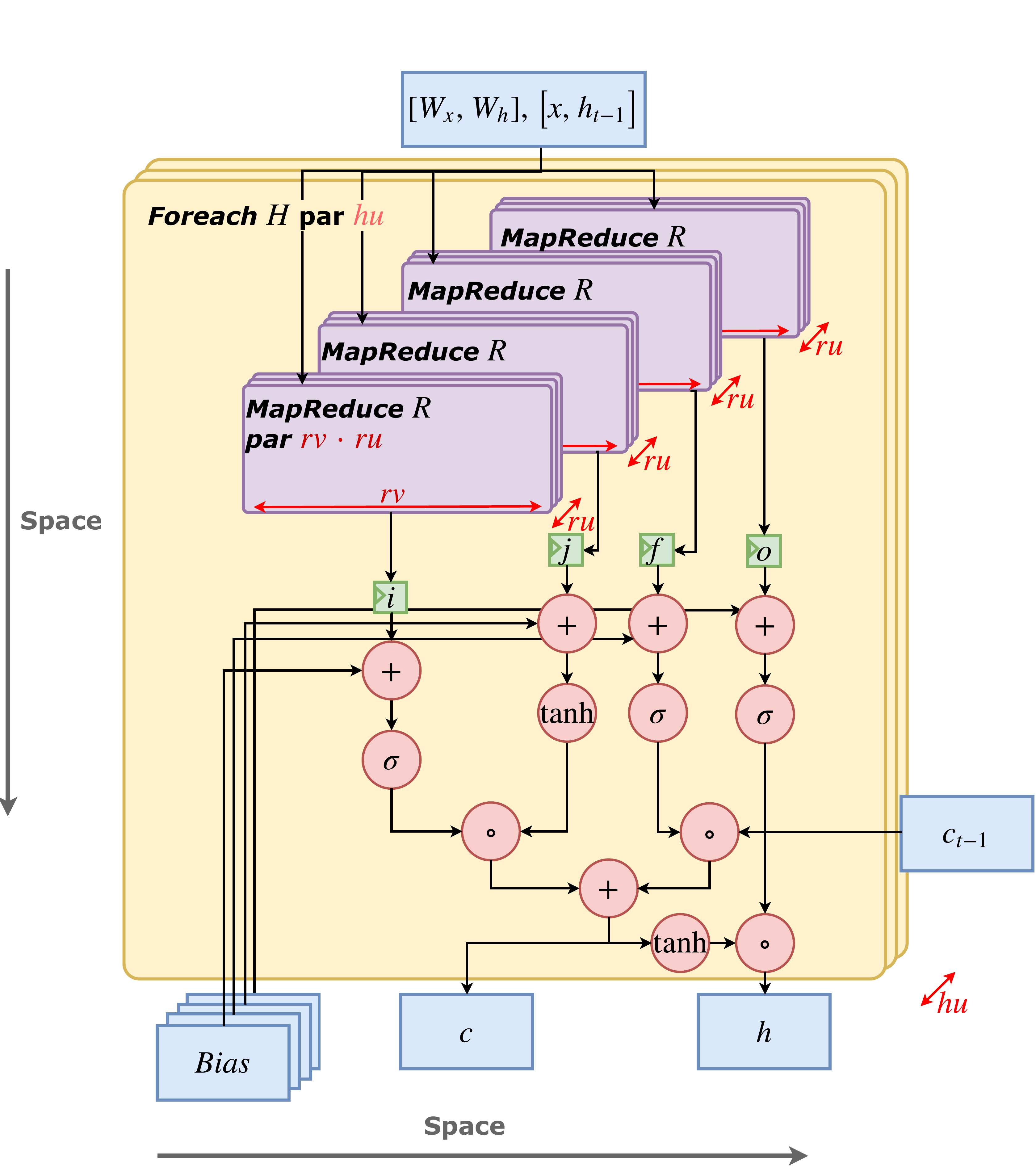}
  \caption{Compute and memory layout of a loop-based LSTM design.}\label{fig:spatial_lstm}
  \vspace*{-0.2in}
\end{figure}
\begin{figure}
  \centering
  \includegraphics[width=\columnwidth]{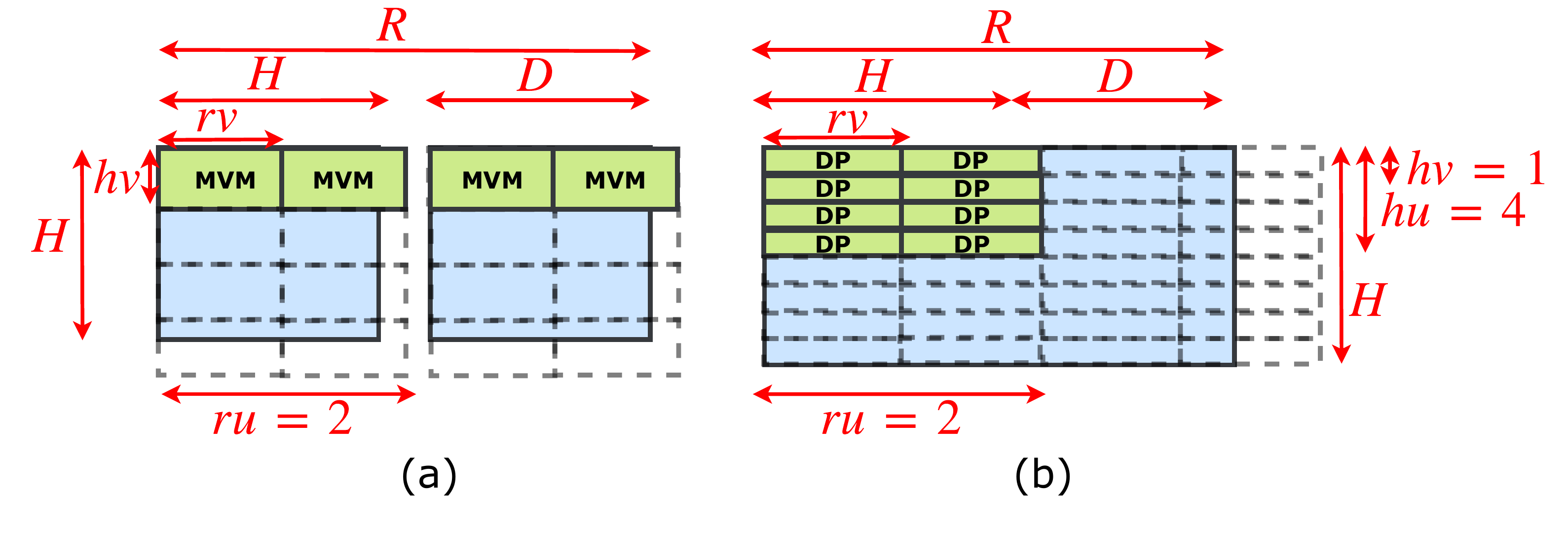}
   \caption{Fragmentation in an MVM-based design (a) and a loop-based design (b) in MVM.}\label{fig:bwt}
\end{figure}

As shown in Figure \ref{fig:spatial_lstm}, each MVM unit is replaced by a MapReduce unit to 
compute the tiled dot product.
Each MapReduce is vectorized by $rv$ with pipelined map function followed by a pipelined reduction tree.
$ru$ is the number of parallel MapReduce units. Results of $ru$ MapReduce blocks are reduced and 
accumulated with another reduction tree (not shown in Figure). Next, the dot product result is passed 
through a chain of function units for executing bias add and non-linear functions. Dot products, bias adds, 
and non-linear functions of the four gates can also be parallelized. Finally, the results of the four gates are pipelined
through a set of function units for element-wise operation in LSTM cell.
At the outer loop, LSTM-1 runs for $\frac{H}{hu}$ iterations,
  where $hu$ is the number of parallel LSTM-1 implementations.

In the loop-based design, all intermediate buffers are scalars as opposed to vectors. 
Regarding utilization, the loop-based LSTM design suffers from less underutilization due to unaligned problem size 
compared to the tiled MVM approach in BW. Figure \ref{fig:bwt} shows sources of such underutilizations.
An MVM approach design would suffer from 2-D fragmentation on both the $H$ and $D$ dimensions (Figure \ref{fig:bwt} (a)), whereas
the loop-based design only suffers from 1-D fragmentation on the $R$ dimension (Figure \ref{fig:bwt} (b)).

\begin{figure}
  \centering
  \newsavebox{\lstm}
  \begin{lrbox}{\lstm}
\begin{lstlisting}[language=Spatial,linewidth=1.0\columnwidth]
// Number of hidden units in h and features in x
val H, D = ...
// Loop unrolling / vectorization parameters
val hu, ru, hv, rv = ...
val c = SRAM[T](H) // SRAM storing C
val xh = SRAM[T](D+H) // SRAM storing [X,H]
// Concatenated weights [Wx,Wh] for each gate in 2-D SRAMs
val wi, wj, wf, wo:SRAM2[T] = ...
val bi, bj, bf, bo: SRAM[T] = ... // Bias
// Lookup tables for non-linear functions
val luti, lutj, luf, luto: SRAM[T] = ...
val tanh:SRAM[T] = ... // Lookup table for tanh
Sequential.Foreach (nSteps by 1){ step =>
  // Loop range from 0 to H parallelized by hu
  Foreach(H par hu){ ih =>
    def fusedDotProductWithNonLinear(w:SRAM2[T], lut:SRAM[T], b:SRAM[T]) = {
      // Tiled dot product with blocking size of rv parallelized by ru
      val elem = Reduce(Reg[T])((D+H) by rv par ru){ iu =>
        Reduce(Reg[T])(rv par rv){ iv =>
          val iuv = iu + iv
          w(ih, iuv) * xh(iuv)
        }{ (a,b) => a + b }
      }{ (a,b) => a + b }.value + b(ih)
      lut(elem)
    }
    val i = fusedDotProductWithNonLinear(wi, luti, bi)
    val j = fusedDotProductWithNonLinear(wj, lutj, bj)
    val f = fusedDotProductWithNonLinear(wf, lutf, bf)
    val o = fusedDotProductWithNonLinear(wo, luto, bo)
    val cNew = i*j + c(ih) * f
    c(ih) = cNew
    xh(ih+D) = tanh(cNew) * o
  }
}
\end{lstlisting}
  \end{lrbox}
  \begin{tabular} {c}
    \usebox{\lstm} \\
  \end{tabular}
  \caption{Example of LSTM in Spatial.}
\label{fig:spatial_app}
\end{figure}

Figure \ref{fig:spatial_app} shows a loop-based LSTM design implemented in Spatial.
\textbf{Foreach} is a loop construct with a lambda that takes loop iterator as input.
\textbf{Reduce} is a construct that executes MapReduce by taking a map function followed
by a reduction function. User declare explicit on-chip scratchpads and registers with
\textbf{SRAM} and \textbf{Reg}.
To enable fine-tuning an RNN application, we exposes loop vectorization factor $rv, hv$ and
unrolling factors $hu, ru$.

  \section{Plasticine Specialization for RNN Serving}
\label{sec:arch}
To show efficient execution of the loop and parallel pattern constructs,
  we map our implementation onto a spatial architecture, Plasticine.
\textbf{Foreach} at Line $17, 19$ and \textbf{Reduce} at Line $22, 23$
  are mapped to PCUs on Plasticine.
When the application size is small,
  these constructs are executed using pipelined SIMD lanes within a single PCU.
When the application size is large,
  multiple PCUs can be used to parallelize and pipeline the dot
  product across PCUs. Element-wise operations can be executed in a deep pipeline
  formed by chaining multiple PCUs.

To fit an RNN's weights on-chip,
  we execute our application with low-precision arithmetics.
In this section,
  we propose the necessary micro-architectural changes to
  support low-precision arithmetics on Plasticine.
We also discuss architectural parameter selection for Plasticine
  to serve RNN applications efficiently.

\subsection{Mixed-Precision Support}
\label{sec:arch:varprec}
Previous works \cite{fowers2018configurable, jouppi2017datacenter}
  have shown that low-precision inference can deliver promising performance
  improvements without sacrificing accuracy.
In the context of reconfigurable architectures such as FPGAs,
  low-precision inference not only increases compute density,
  but also reduces required on-chip capacity for
  storing weights and intermediate data.

To support low-precision arithmetics without sacrificing coarse-grained reconfigurability,
we introduce two low-precision struct types in Spatial: a tuple of 4 8-bit and 2 16-bit floating-point 
numbers, \texttt{4-float8} and \texttt{2-float16} respectively.
Both types packs multiple low-precision values into a single precision storage.
We support only 8 and 16-bit precisions, which are commonly seen in deep learning inference hardwares.
Users can only access values that are 32-bit aligned.
This constraint guarantees that the microarchitectual change is only local to the PCU.
Banking and DRAM access granularity remains intact from the original design.

\begin{figure}
  \centering
  \includegraphics[width=1\columnwidth]{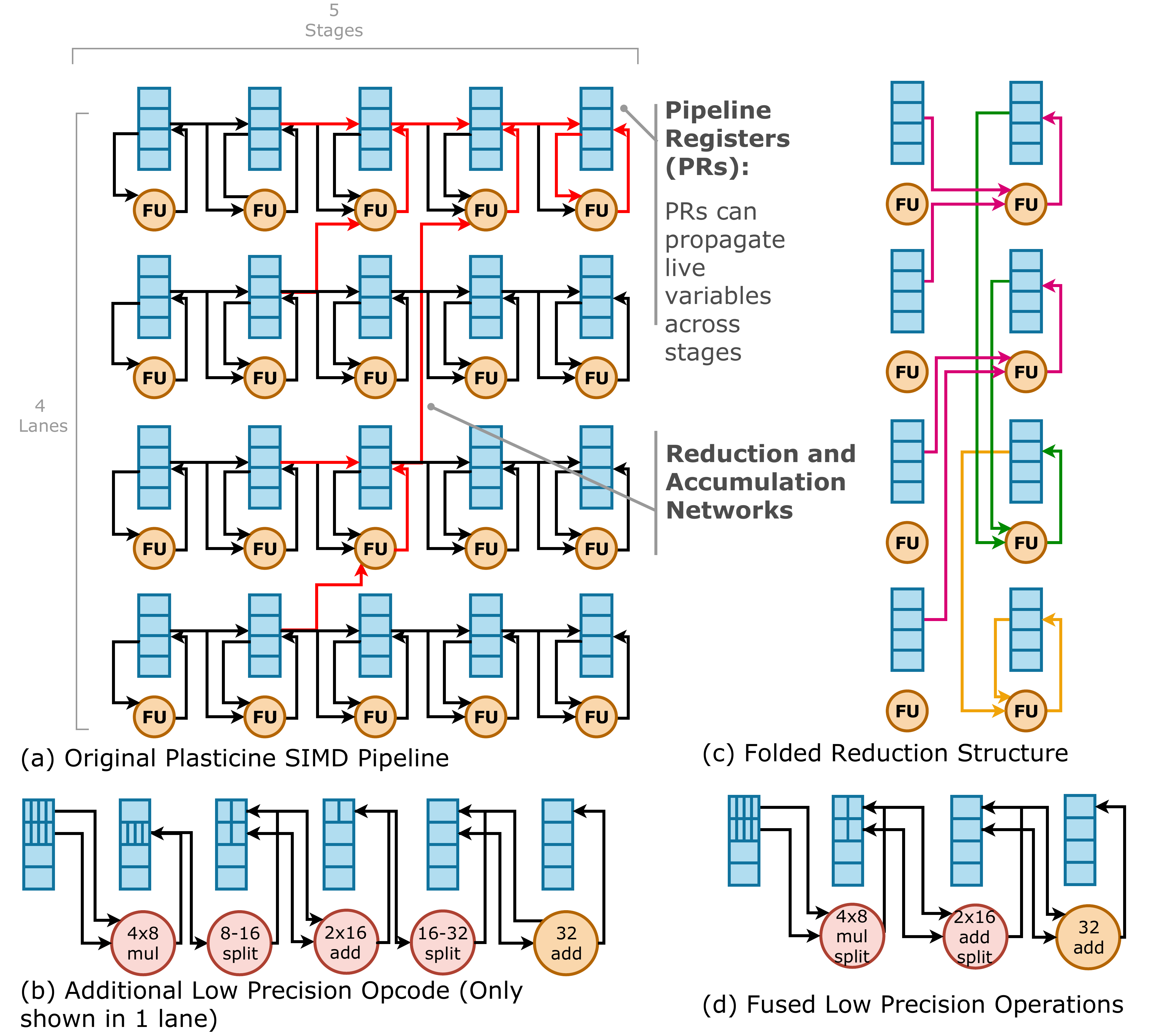}
  \caption{Plasticine PCU SIMD pipeline and low-precision support.
  Red circles are the new operations. Yellow circles are the original opertaions in Plasticine.
  In (d) the first stage is fused $1^{st}, 2^{nd}$ stages, and the second stage is fused
  $3^{nd}$, $4^{th}$ stages of (b).
   }
  \label{fig:lowprec}
  \vspace*{-0.3in}
\end{figure}
Figure \ref{fig:lowprec} (a) shows the original SIMD pipeline in a Plasticine PCU.
Each FU supports both floating-point and fix-point operations.
When mapping applications on Plasticine,
  the inner most loop body is vectorized across the lanes of the
SIMD pipeline, and different operations of the loop body are mapped to different stages.
Each pipeline stage contains a few pipeline registers (PRs)
  that allow propagation of live variables across stages.
Special cross-lane connections as shown in red in Figure \ref{fig:lowprec} enable reduction operations.
To support 8-bit element-wise multiplication and 16-bit reduction, we add 4 opcodes to the FU, shown in
Figure \ref{fig:lowprec} (b).
The $1^{st}$ and $3^{rd}$ stages are element-wise, low-precision operations
  that multiply and add 4 8-bit and 2 16-bit values, respectively.
The $2^{nd}$ and $4^{th}$ stages rearrange low-precision values into two registers,
  and then pad them to higher precisions.
The $5^{th}$ stage reduces the two 32-bit value to a single 32-bit value using the existing add operation. 
From here, we can use the original
reduction network shown in Figure \ref{fig:lowprec} (a) to complete the remaining reduction and accumulates
in 32-bit connection.

With 4 lanes and 5 stages,
  a PCU first reads 16 8-bit values,
  performs 8-bit multiplication followed by rearrangement and padding,
  and then produce 16 16-bit values after the second stage.
The intermediate values are stored in 2 PRs per lane.
Next, 16 16-bit values are reduced to 8 16-bit values
  and then rearranged to 8 32-bit value in 2 PRs per lane.
Then, the element-wise addition in 32-bit value
  reduces the two registers in each line into 4 32-bit values.
These values are fed through the
  reduction network that completes the remaining
  reduction and accumulation in two plus one stages.

In a more aggressive specialization,
  we can fuse the multiply and rearange into the same stage.
We also fuse the first low-precision reduction with the next rearange as shown in Figure \ref{fig:lowprec} (d).
In this way, we can perform the entire low-precision map-reduce in 2 stages
  in addition to the original full precision reduction.
In order to maximize hardware reuse,
  we assume that it is possible to construct a full precision FU
  using low-precision FUs.
In addition, we observe that the original reduction network in the SIMD lanes
  could lead to low FU utilization.
To improve FU utilization, we fold the entire tree structure in a single stage.
Figure \ref{fig:lowprec} (c) shows the folded reduction accumulation structure.
Specifically, latter reductions in the tree are mapped to earlier stages in the pipeline.
In this setup, the entire reduction plus accumulation
  is still fully pipelined in $\log_2(\#_{LANE})+1$ cycles
  with no structural hazard.
With fused reduced-precision multiplication and reduction, and folded reduction tree,
  a PCU is able to perform all map-reduce that accumulates $4 \#_{LANE}$
  8-bit values using 4 stages.
All the operations are completed in $2+\log_2(\#_{LANE})+1$ cycles.

\subsection{Sizing Plasticine for Serving RNN}
\begin{figure}
  \centering
  \includegraphics[width=\columnwidth]{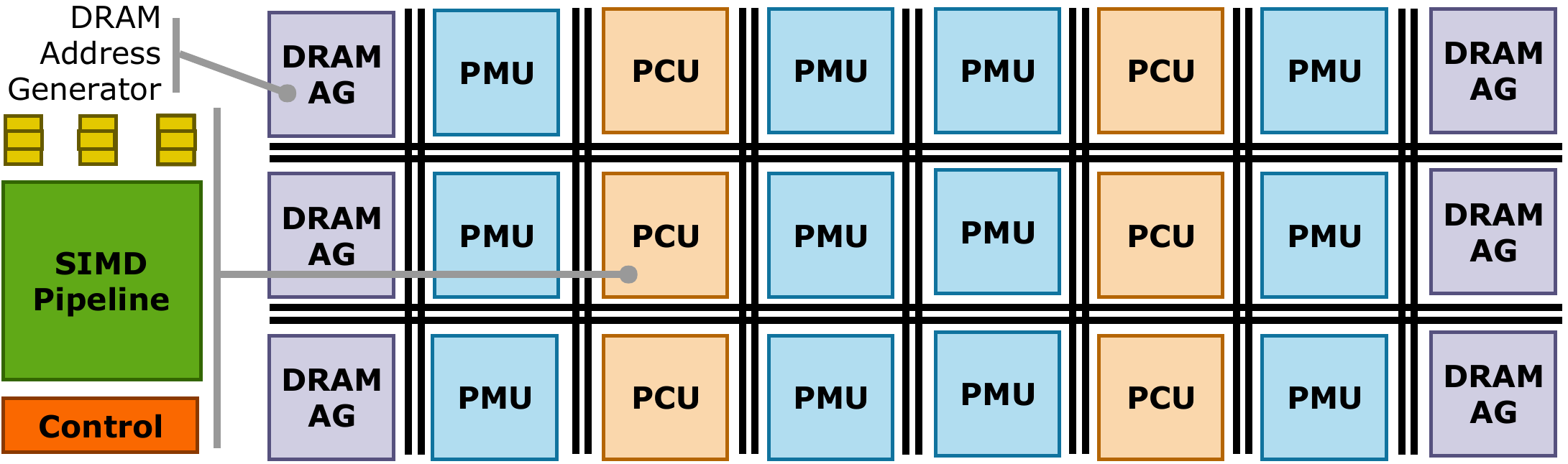}
  \caption{Variant configuration of Plasticine for serving RNN.}
  \label{fig:arch}
  \vspace*{-0.2in}
\end{figure}

Evaluating an RNN cell containing $N$ hidden units and $N$ input features
  requires $2N^2$ computations and $N^2+N$ memory reads.
With large $N$, the compute to memory ratio is 2:1.
The original Plasticine architecture uses a checkerboard layout
  with 1 to 1 ratio between PCU and PMU.
A PCU has 6 stages and 16 lanes, and a PMU has 16 banks.
This provides a 6:1 ratio between
  compute resource and on-chip memory read bandwidth.
As a result of this layout,
  on-chip memory read bandwidth becomes the bottleneck for accelerating RNN serving applications.
Given that RNNs cover a wide range of important applications,
  we select a Plasticine configuration tailored for RNN serving.
Specifically, we choose a 2 to 1 PMU-PCU ratio with 4 stages in each PCU.
Figure \ref{fig:arch} shows the layout of this Plasticine variant.
  \section{Evaluation} \label{sec:eval}
In this section, we evaluate the real-time RNN serving tasks on various platforms.
We start with the methodology of our experiments, followed by a discussion of performance and power comparisons across
these platforms.

\subsection{Methodology} \label{sec:methodology}
To evaluate RNN serving, we use the LSTM and GRU tasks from Baidu DeepBench as our benchmarks.
We evaluate the benchmarks across processor-based architectures including CPU and GPU, 
and spatial architectures including FPGA and CGRA.
Table \ref{tab:spec} shows the detailed specifications of the targeting hardware, 
which includes state-of-the-art high performance platforms in each of the commercialized categories.
Table \ref{tab:appconf} summarizes application configurations of each platform.

\paragraph{CPU} We implement the applications in TensorFlow 1.10, and evaluate our implementations on 
Intel Xeon Scalable Processor (Skylake) CPU.
We use the \texttt{LSTMBlockFusedCell} and \texttt{GRUBlockCell} kernels in TensorFlow.
We further enable AVX2 vector instructions for CPU evaluation. Due to lack of low-precision
support in both tool chain and platform, we use single-precision for our implementation.

\paragraph{GPU} We use TensorFlow with cuDNN Library to target NVIDIA Tesla V100 GPU from Google Cloud. 
cuDNN is a GPU-accelerator Library from NVIDIA that is specialized for deep learning.
We use 16-bit precision for our implementation on GPU.
On both CPU and GPU platforms, we run \emph{TensorFlow} profilers and collect the time spent 
only on evaluating the RNN cells.

\paragraph{Plasticine} We implement the applications in Spatial, which targets Plasticine.
Although Spatial has FPGA back-end support, Stratix 10 is not commercially available at the time of the submission of this work.
The current FPGA targets that Spatial support are not comparable to Stratix 10 both in terms of memory and compute capacity. 
Therefore, we only use Spatial to target Plasticine for this evaluation. However, our approach should generally benefit
an implementation on a high performance FPGA like Stratix 10.
We choose Plasticine configuration that matches the peak 8-bit FLOPS and
on-chip scratchpad capacity of a Stratix 10 FPGA. The exact configuration of Plasticine is shown in Table \ref{tab:conf}.
In order to minimize the overhead of low-precision support, Plasticine only supports 8-bit, 16-bit, and 32-bit element-wise 
operations, and mixed precision reduction operation. 
For our evaluation, the element-wise operations are performed in 8-bit precision, 
the first stage of the reduction is performed in 16-bit, 
while the remaining of the reduction and accumulation are performed in 32 bit operations.

To measure the performance, we use a cycle accurate simulator for Plasticine. 
We modified the simulator to model the proposed micro-architectural changes to support low-precision operations.
We use the area and power of individual CUs and network switches from the original Plasticine paper, 
and compute total area of configuration shown in Table \ref{tab:conf}. 
As discussed in Section \ref{sec:arch}, we reduce the number of stages in PCU from 6 stages to 4 stages with fused low-precision
operations and folded reduction tree. 
Low preicision function units can be used to compose full precision units. 
We conservatively estimate the area and power of PCU stays the same with our proposed change and reduced two stages. 
We also increase the PMU to PCU ratio to better match the compute to memory
ratio for RNN inference applications. To match the memory capacity of Stratix 10, we shrink the scratchpad capacity of 
each PMU from 256kB to 84kB.
For power calculations, we generate activity tracing of the CUs from simulation, and then integrate 
with characterized power of individual PCU to compute the total power. The power and area characterizations are based off
synthesis at 28nm technology at 1GHz clock frequency.

\paragraph{Brainwave} Finally, we also compared our results to Microsoft Brainwave framework.
For this evaluation, we compare to Brainwave implemented
on top of Intel Stratix 10 FPGA. Brainwave is synthesized at 250MHz and all operations are performed in
blocked low-precision floating-point format described in section~\ref{sec:blaslstm}.
\begin{table}[t]
\caption{Plasticine configuration.}
\label{tab:conf}
\centering
\scriptsize
\begin{tabular}{L{3cm}rL{2.5cm}r}
\toprule
\# Row                     & 24   & \# Column        & 24  \\
\# PCU                     & 192  & \# PMU           & 384 \\
\# Lanes in PCU            & 16   & \# Stages in PCU & 4   \\
Scrachpad capacity per PMU & 84kB &                  &     \\
\bottomrule
\end{tabular}
\end{table}

\begin{table}
\caption{Hardware specifications for target platforms.}
\label{tab:spec}
\scriptsize
\centering
\begin{tabular}{L{2.5cm}M{1.2cm}M{0.8cm}M{0.8cm}M{1cm}}
\toprule
  Specification        & Intel Xeon Skylake (Dual core) & Tesla V100 SXM2 & Stratix 10 280 FPGA & Plasticine\\
\midrule
Max Clock Rate (GHz) & 2.0/2.8*                  & 1.38/1.53*      & 1                   & 1 \\
On-chip memory** (MB) & 55                        & 20              & 30.5                & 31.5\\
Peak 32-bit TFLOPS   & --                      & 15.7            & 10                  & 12.5\\
Peak 8-bit TFLOPS    & --                        & --              & 48                  & 49\\
Technology ($nm$)    & 14                        & 12              & 14                  & 28\\
Die Area ($mm^2$)    & 64.4                      & 815             & 1200                & 494.37 \\
  TDP (W)    & 15                      & 300             & 148                & 160 \\
\bottomrule
\end{tabular}
* Base/Boosted Frequency
** Capacity of L3 cache for CPU, register file for GPU, and on-chip scratchpad for reconfigurable architectures.
\end{table}

\begin{table}
\caption{Application configurations for target platforms.}
\label{tab:appconf}
\centering
\scriptsize
\begin{tabular}{L{1.8cm}M{1cm}M{1cm}M{1cm}M{1.5cm}}
\toprule
Platform                       & Intel Xeon Skylake & Tesla V100 SXM2 & Stratix 10 280 FPGA & Plasticine\\
\midrule
Software Framework             & TF+AVX2                   & TF+cuDNN        & Brainwave           & Spatial \\
Achieved Clock Frequency (GHz) & 2                         & 1.38            & 0.25                & 1 \\
Precision                      & f32                       & f16             & blocked precision   & mix f8+16+32\\
\bottomrule
\end{tabular}
\end{table}

\begin{table*}
\caption{Performance comparison of DeepBench Inference.}
\label{tab:eval}
\centering
\scriptsize

  \begin{tabular}{|L{0.6cm}|M{0.4cm}|M{0.4cm}|M{1.1cm}M{0.45cm}M{0.45cm}M{1.2cm}|M{1.1cm}M{0.45cm}M{0.45cm}M{1.2cm}|M{1.1cm}M{0.6cm}M{0.6cm}|M{1.2cm}|}
\hline
    \multicolumn{3}{|c|}{\sc Benchmarks}					&	\multicolumn{4}{c|}{\textsc{Latency} (ms)}							&	\multicolumn{4}{c|}{\sc Effective TFLOPS}							&	\multicolumn{3}{M{3cm}|}{\sc Plasticine Speedup (x)} & \sc Power (W)			\\ \hline
    &	\sc H	&	\sc T	&	\sc Xeon Skylake	&	\sc Tesla V100	&	\sc BW &	\sc Plasticine	&	\sc Xeon Skylake	& Tesla V100	&	\sc BW	&	\sc Plasticine	&	\sc Xeon Skylake	&	\sc Tesla V100	&	\sc BW	 & \sc Plasticine\\ \hline
\multirow{5}{*}{\sc\bf LSTM}	&	256	&	150	&	15.75	&	1.69	&	0.425	&	0.0419	&	0.010	&	0.09	&	0.37	&	3.8	&	376.3	&	40.4	&	10.2	&	28.5	\\
	&	512	&	25	&	11.50	&	0.60	&	0.077	&	0.0139	&	0.009	&	0.18	&	1.37	&	7.6	&	830.3	&	43.2	&	5.6	&	53.7	\\
	&	1024	&	25	&	107.65	&	0.71	&	0.074	&	0.0292	&	0.004	&	0.59	&	5.68	&	14.4	&	3,686.6	&	24.3	&	2.5	&	97.2	\\
	&	1536	&	50	&	411.00	&	4.38	&	0.145	&	0.1224	&	0.005	&	0.43	&	13.01	&	15.4	&	3,357.8	&	35.8	&	1.2	&	102.7	\\
	&	2048	&	25	&	429.36	&	1.55	&	0.074	&	0.1060	&	0.004	&	1.08	&	22.62	&	15.8	&	4,050.6	&	14.6	&	0.7	&	104.5	\\ \hline
\multirow{6}{*}{\sc\bf GRU}	&	512	&	1	&	0.91	&	0.39	&	0.013	&	0.0004	&	0.003	&	0.01	&	0.25	&	7.6	&	2,182.3	&	942.4	&	31.2	&	61.9	\\
  &	1024	&	1500	&	3,810.00	&	33.77	&	3.792	&	1.4430	&	0.005	&	0.56	&	4.98	&	13.1	&	2,640.3	&	23.4	&	2.6	&	109.1	\\
  &	1536	&	375	&	2,730.00	&	13.12	&	0.951	&	0.7463	&	0.004	&	0.81	&	11.17	&	14.2	&	3,658.3	&	17.6	&	1.3	&	114.6	\\
	&	2048	&	375	&	5,040.00	&	17.70	&	0.954	&	1.2833	&	0.004	&	1.07	&	19.79	&	14.7	&	3,927.5	&	13.8	&	0.7	&	101.2	\\
	&	2560	&	375	&	7,590.00	&	23.57	&	0.993	&	1.9733	&	0.004	&	1.25	&	29.69	&	15.0	&	3,846.4	&	11.9	&	0.5	&	117.2	\\ \hline
\multicolumn{3}{|c|}{\textsc{\bf Geometric Mean}}					&		&		&		&		&		&		&		&		&	2,529.3	&	29.8	&	2.0	&		\\
\hline
\end{tabular}
\end{table*}

\begin{table}
\caption{Loop unrolling and vectorization parameters for spatial architectures.}
\label{tab:param}
\centering
\scriptsize
\begin{tabular}{|L{0.5cm}|M{0.4cm}|M{0.4cm}|M{0.3cm}|M{0.3cm}|M{0.3cm}|M{0.3cm}|M{0.3cm}|M{0.3cm}|M{0.3cm}|}
\hline
  \multicolumn{3}{|c|}{\sc Benchmarks}						&\multicolumn{3}{c|}{\sc Stratix 9 BW} &						\multicolumn{4}{c|}{\sc Plasticine}							\\\hline
	&	\sc H	&	\sc T	&	$ru$	&	$hv$	&	$rv$	&	$hu$	&	$hv$	&	$ru$	&	$rv$	\\\hline
\multirow{5}{*}{\sc\bf LSTM}	&	256	&	150	&	\multirow{11}{*}{6}	&	\multirow{11}{*}{400}	&	\multirow{11}{*}{40}	&	6	&	\multirow{11}{*}{1}	&	4	&	\multirow{11}{*}{64}	\\ \cline{7-7} \cline{9-9}	\cline{2-3}
	&	512	&	25	&		&		&		&	\multirow{4}{*}{4}	&		&	\multirow{10}{*}{8}	&		\\	\cline{2-3}
	&	1024	&	25	&		&		&		&		&		&		&		\\	\cline{2-3}
	&	1536	&	50	&		&		&		&		&		&		&		\\	\cline{2-3}
	&	2048	&	25	&		&		&		&		&		&		&		\\ \cline{7-7}	\cline{2-3}
\multirow{6}{*}{\sc\bf GRU}	&	512	&	1	&		&		&		&	\multirow{6}{*}{2}	&		&		&		\\	\cline{2-3}
	&	1024	&	1500	&		&		&		&		&		&		&		\\	\cline{2-3}
	&	1536	&	375	&		&		&		&		&		&		&		\\	\cline{2-3}
	&	2048	&	375	&		&		&		&		&		&		&		\\	\cline{2-3}
	&	2560	&	375	&		&		&		&		&		&		&		\\	\cline{2-3}
	&	2816	&	750	&		&		&		&		&		&		&		\\\hline	\cline{2-3}
\end{tabular}
\end{table}

\subsection{RNN Performance Analysis} \label{sec:results}
Table \ref{tab:eval} shows the performance comparison of LSTM and GRU with various numbers of hidden units (H) and step sizes (T) over
the four platforms. Overall, both CPU and GPU significantly underutilize the available compute FLOPS.
In addition, they cannot meet the latency requirement for real-time serving for all problem sizes.
Both BW and Plasticine deliver promising latencies within 5ms for all problem sizes.
When serving very large RNNs, BW provides better performance
	with up to 2x better than Plasticine on the largest GRU (H=2816).
When serving small and medium size RNNs, Plasticine performs better than BW
	with up to 30x better performance on small GRU (H=512).
We also observe that Plasticine delivers consistent FLOPS when serving all the problem sizes.

\paragraph{Processor-Based Architectures}
For CPU experiments, the RNN kernels from TensorFlow itself is not multi-threaded.
Since we focus on real-time serving of RNN applications, we use batch size of 1 for all of our benchmarks,
	which expose no parallelism outside the kernel level.
Consequently, the machine is still very underutilized even with AVX2 instruction.
Although one could implement RNN directly in c++,
	the MVM sizes in RNNs are too small to benefit from multi-threading due to the synchronization overhead.
V100 with cuDNN library provides significant acceleration compared to CPU.
	Nevertheless, the latency is still high.
This is because GPUs are designed for throughput oriented rather than latency sensitive workloads.
Provided that the library is based on BLAS3 routines, which are matrix-matrix operation, MVMs in 
RNN serving suffer from significant resource underutilization.
In Table \ref{tab:eval}, V100 shows very poor performance on GRU (H=512). This is likely due to
the initialization overhead which should not be timed.
From our evaluation, neither processor-based architectures are suitable for providing low-latency serving on
RNN applications.

\paragraph{Spatial Architectures} Table \ref{tab:param} shows the selected design parameters for each 
problem size for BW and Plasticine.
On Stratix 10, BW uses 6 tile engines ($ru$) with native dimension of 400 ($hv$) and 40 lanes ($rv$).
Large $hv$ and $rv$ improve the data-to-control ratio by amortizing the scheduling overhead over a large vectorized instruction.
However, this design choice aggravates the underutilization for small RNN feature sizes at 256 and 512.
Our implementation effectively uses $hv$ of size 1 by performing dot product instead of MVM, 
which prevents fragmentation in the $H$ dimension.
	With $hv=1$, all the intermediate buffers are stored in registers.
In contrast, BW uses register files of size $hv$.
In addition, our proposed implementation captures additional gate-level, X, and H parallelism as well as pipelining at element-wise functions.
In contrast, BW schedules these operations in time and dispatches corresponding instructions to drive the compute units.

A CGRA is less flexible than an FPGA when performing arbitrary low-precision operations. 
In this example,
	we increase memory density of Plasticine by supporting quantile precisions as described in Section \ref{sec:arch:varprec}.
All weights are stored in 8 bit format, so as the multiplication operations of MVM. 
The reduction and accumulation operations are implemented in mix of 16 and 32 bit precisions.
Hence, the peak FLOPS when performing mixed precision map-reduce is much less than the peak FLOPS for blocked low-precision format in BW.
As a result, Plasticine performs worse than BW on the large RNNs.

In addition, Plasticine delivers very consistent FLOPS for
different problem sizes. For small problem size, the dot product can be fully unrolled with $rv * ru$. Therefore, we can
increase $hu$ to explore additional parallelism across the hidden units. For large problem size, dot product becomes the bottleneck of
the pipeline. Hence, we reduce $hu$ and increase $ru$ to balance the throughput between dot product and element-wise operations.
In this example, BW uses a single set of parameters for all problem sizes.
Although one can potentially tune parameters for different problem sizes,
	doing so will incur re-synthesis and place-and-route on an FPGA,
	which is an order of magnitude longer than the compilation time needed for a CGRA design.
In addition, to exhaust hardware resources with a smaller $hv$,
	one would have to increase the number of matrix vector tile engines $hu\times ru$ in BW.
As a result, decoders and schedulers associated with these units
	will drive up the control-to-data overhead and deliver less FLOPS for larger problem sizes.

\subsection{Area and Power Analysis} \label{sec:results}
Table \ref{tab:spec} shows the die area comparison of different platforms.
	While the GPU has a publicly-reported die area measurement \cite{markidis2018nvidia},
	Xeon Skylake and Stratix 10 only have estimated die areas based on
	their estimated transistor counts \cite{inteldie}.
With the rough area estimates, we can see that while CPU has the smallest area in this case,
	the performance gap is too large even after we scale up to a 28-core server.
The GPU also delivers bad performance per area mostly due to the low utilization of compute FLOPS.
Stratix 10 delivers the best performance for the large RNNs,
	but with the largest die area estimates of 30 billion transistors \cite{stratix10die}.
Plasticine's die area is based on the synthesis results at 28nm,
	which is one generation older than all the other platforms.
With technology scaling,
	Plasticine should possess double the amount of compute and memory resources at 14nm for the same die area,
	which will roughly match Stratix 10's performance on all the RNN problem sizes.
At the same time, Plasticine is more than 2x smaller than Stratix 10,
	which could also contribute at least 2x - 60x performance per area improvement for all problem sizes.
Table \ref{tab:spec} shows the thermal design power (TDP) of the four platforms,
	which is the peak power achievable for any workloads \cite{inteltdp, stratix10tdp, v100spec}.
BW also reports a measured peak power for the given set of benchmarks of 125W.
Table \ref{tab:eval} shows the simulated power for Plasticine for each benchmark.
	Overall, the peak power among benchmarks for Plasticine is 118W,
	which is slightly less than the peak power compared to BW.

  \section{Related Work}
\label{sec:related}

Previously proposed serving platforms
  focus on exploiting data locality by mapping RNN cells onto spatial architectures.
For example, Chang et al presented an FPGA-based implementation of an LSTM network \cite{chang2015recurrent}.
This approach works well for supporting small RNNs.
However, for a large RNN, the weights would be too large to fit on-chip.
As a result, the serving latency would be dominated by DRAM data loading.
To address the issue of fitting RNN weights on-chip,
  several previous works \cite{han2016dsd, wang2018c, see2016compression, narang2017exploring}
  have studied the approaches for compressing RNN weights.
For example, Han et al presented a compression scheme called DSD \cite{han2016dsd}.
  It iteratively removes parameters in the weight matrices
  and retrains the sparse model
  to minimize the accuracy loss introduced by sparsity \cite{han2016dsd}.
With this compression scheme,
  Han et al were able to deploy an LSTM network
  containing 3.2 million parameters onto a modern FPGA
  without sacrificing accuracy.
Compared to serving on CPU and GPU platforms,
  serving a sparse LSTM network on FPGA provides
  much lower latency and higher energy efficiency.
However, we find that it could be hard to generalize
  this compression scheme for all the RNN tasks.
RNNs are very flexible in terms of their model structures.
Applying a DSD-like compression scheme to all the RNN models
  requires hand-tuning the compression heuristics for every model.
To avoid hand-tuning,
  He et al proposed an approach that uses reinforcement learning
  techniques for automatic compression tuning \cite{he2018amc}.
However, their approach focuses on compressing CNN tasks on edge devices,
  which may not be transferrable to the case of serving RNN tasks in datacenter.
Observing that the sparsity-based compression schemes are still under active development,
  we choose to support compression schemes that focus on representing RNN weights
  using low-precision data format.
Commercially available platforms
  such as Google TPU \cite{jouppi2017datacenter}
  and Microsoft BrainWave \cite{fowers2018configurable}
  support these schemes.


  \section{Conclusion}
\label{sec:conclusion}

In this paper, we describe a set of techniques for performing cross-kernel optimization within RNN cells.
We identify that by moving away from BLAS abstraction and focus on optimizing loop-level construct,
we are able to achieve consistent hardware utilization when serving RNN cells of different sizes.
We show that we are able to achieve 10-20x performance improvement at a less advanced technology
compared to the state-of-the-art GPU platform, and a geometric speedup of 2x compared to the state-of-the-art FPGA-based platform.
  \section{Acknowledgement}

We appreciate the anonymous reviewers for their feedback.
We thank Matthew Feldman for compiler support
  and his constructive suggestions on the manuscript of this paper,
  and Raghu Prabhakar for providing insights and
  feedback on the architecture section of this paper.
We also thank Google for the cloud credits.
This material is based on research sponsored by Air Force Research Laboratory (AFRL) and Defense
Advanced Research Projects Agency (DARPA) under agreement number FA8650-18-2-7865. The U.S.
Government is authorized to reproduce and distribute reprints for Governmental purposes
notwithstanding any copyright notation thereon. The views and conclusions contained herein are those
of the authors and should not be interpreted as necessarily representing the official policies or
endorsements, either expressed or implied, of Air Force Research Laboratory (AFRL) and Defense
Advanced Research Projects Agency (DARPA) or the U.S. Government.
This research is also supported in part by affiliate members and other supporters of the
Stanford DAWN project - Ant Financial, Facebook, Google, Infosys, Intel, Microsoft, NEC, Teradata,
SAP and VMware.

  \bibliographystyle{sysml2019}
  \bibliography{references}
\end{document}